\documentclass[aps,pra,twocolumn,showpacs,10pt]{revtex4-1}
\usepackage{amsmath,amssymb,graphicx,stmaryrd}

\usepackage[utf8]{inputenc}
\usepackage[T1]{fontenc}
\usepackage{abraces}

\IfFileExists{newtxtext.sty}
   {\usepackage{newtxtext,newtxmath}}
   {\IfFileExists{stix.sty}
      {\usepackage{stix}}
      {\IfFileExists{mathptmx.sty}
      {\usepackage{mathptmx}}{} } }

\usepackage{textcomp}

\usepackage{bm}
\usepackage{bbold}

\IfFileExists{siunitx.sty}{\usepackage{booktabs,siunitx}}{}

\renewcommand{\vec}[1]{\bm{#1}}
\newcommand{\bra}[1]{\langle#1\rvert}
\newcommand{\ket}[1]{\lvert#1\rangle}

\newcommand{\expectation}[1]{\langle#1\rangle}

\newcommand{\linkdoi}[2]{\href{http://dx.doi.org/#2}{#1}}
\newcommand{\linkarxiv}[2]{\href{http://arxiv.org/abs/#2}{#1}}
\newcommand{\linkisbn}[2]{\href{https://isbndb.com/book/#2}{#1}}
\newcommand{\linkapsmeeting}[2]{\href{http://meetings.aps.org/Meeting/#2}
{#1}}
\newcommand{\linkhttp}[2]{\href{http://#2}{#1}}

\pdfoutput=1
\usepackage{color}
\definecolor{LinkColor}{rgb}{0.256,0.439,0.588}
\usepackage{hyperref}
\hypersetup{
   pdfauthor={Khagendra Adhikari and K. S. D. Beach},
   pdftitle={Deforming the Fredkin spin chain away from its frustration-free point},
   pdfsubject={ED and DMRG study of the Fredkin spin chain},
   pdfkeywords={Fredkin,} {quantum magnetism,} {phase transitions,} {competing ordered phases,} {ferromagnetism,} {antiferromagnetism,} {dimerization},
   colorlinks=true,
   citecolor=LinkColor,
   linkcolor=LinkColor,
   urlcolor=LinkColor
   }

\begin{document}

\title{Deforming the Fredkin spin chain away from its frustration-free point}

\author{Khagendra Adhikari}
\email[Electronic address:\ ]{kadhikar@go.olemiss.edu}
\affiliation{Department of Physics and Astronomy, The University of Mississippi, University, Mississippi 38677, USA}

\author{K. S. D. Beach}
\email[Electronic address:\ ]{kbeach@olemiss.edu}
\affiliation{Department of Physics and Astronomy, The University of Mississippi, University, Mississippi 38677, USA}

\begin{abstract}
The Fredkin model describes a spin-half chain segment subject to three-body, correlated-exchange interactions and twisted boundary conditions. The model is frustration-free, and its ground state wave function is known exactly. Its low-energy physics is that of a strong xy ferromagnet with gapless excitations and an unusually large dynamical exponent. We study a generalized spin chain model that includes the Fredkin model as a special tuning point and otherwise interpolates between the conventional ferromagnetic and antiferromagnetic quantum Heisenberg models. We solve for the low-lying states, using exact diagonalization and density-matrix renormalization group calculations, in order to track the properties of the system as it is tuned away from the Fredkin point; we also present exact analytical results that hold right at the Fredkin point.  We identify a zero-temperature phase diagram with multiple transitions and unexpected ordered phases. The Fredkin ground state turns out to be particularly brittle, unstable to even infinitesimal antiferromagnetic frustration. We remark on the existence of an ``anti-Fredkin'' point at which all the contributing spin configurations have a spin structure exactly opposite to those in the Fredkin ground state.
\end{abstract}

\maketitle

\section{Introduction}
\label{SECT:Introduction}

The construction of a spin-1 quantum spin chain 
with an exactly solvable ground state~\cite{Bravyi-PRL-12} 
and its later generalization to spin-$S$ (for all integer $S > 1$)
by Movassagh and Shor~\cite{Movassagh-PNAS-16}
have proved to be incredibly fruitful developments.
These models describe quantum spins interacting
locally in the form of a projector, and they 
are frustration-free in the sense that it is possible to 
minimize each term in the Hamiltonian individually.
One finds that the unique ground state wave function is an equal-weight
superposition of all Motzkin walks~\cite{Donaghey-JCT-77} (colored for $S>1$).
The resulting structure is similar in spirit to 
that of the Rokhsar-Kivelson point in the quantum dimer model~\cite{Kivelson-PRB-87,Rokhsar-PRL-88}, where all 
short-range dimer tilings contribute equally in the ground state.

Although the ground state has this exact form, 
the energy spectrum and other properties are less well known.
Analytical work~\cite{Movassagh-PNAS-16,Movassagh-JMP-17} points to some very interesting and unusual behavior. Strong mathematical bounds have been placed on the scaling of the 
excitation gap and the bipartite entanglement entropy as a function
of system size, $N$. 
The first surprise is that the system is gapless, as opposed to the more typical situation for exactly solvable spins chains~\cite{Majumdar-JMP-69,Affleck-PRL-87}.
Second, the excitation gap vanishes polynomially with a dynamical
exponent $z>2$, which implies that these systems cannot be described at low energy by any conformal field theory.
(Numerical estimates of the dynamical exponent range from 2.7 to 3.2~\cite{Bravyi-PRL-12,DellAnna-PRB-16,Chen-PRB-17,Chen-JPA-17}.)
Third, the entanglement entropy shows violation of the area law~\cite{Eisert-RMP-10}, growing as fast as $\sqrt{N}$.

It was once part of the lore of the field that a unique ground state of a physically plausible (local, translationally invariant) one-dimensional Hamiltonian might violate area law scaling by at most a logarithmic correction. The family of Motzkin spin 
chains provides an important counterexample.
Indeed, systematic frustration-free deformations of the model~\cite{Zhang-PNAS-17,Zhang-JPA-17}
yield a tunable family of Motzkin-walk wave functions (no longer equally weighted)
that show a transition between an area-law-compliant phase and 
an area-law-violating phase in which the entanglement entropy is extensively large
(${\sim}N$, with maximal violation corresponding to the ``rainbow'' state
proposed in Ref.~\citenum{Ramirez-JSM-15}) and the energy gap 
vanishes exponentially~\cite{Levine-JPA-17}.

\begin{figure}
\begin{center}
\includegraphics[width = 0.85\columnwidth]{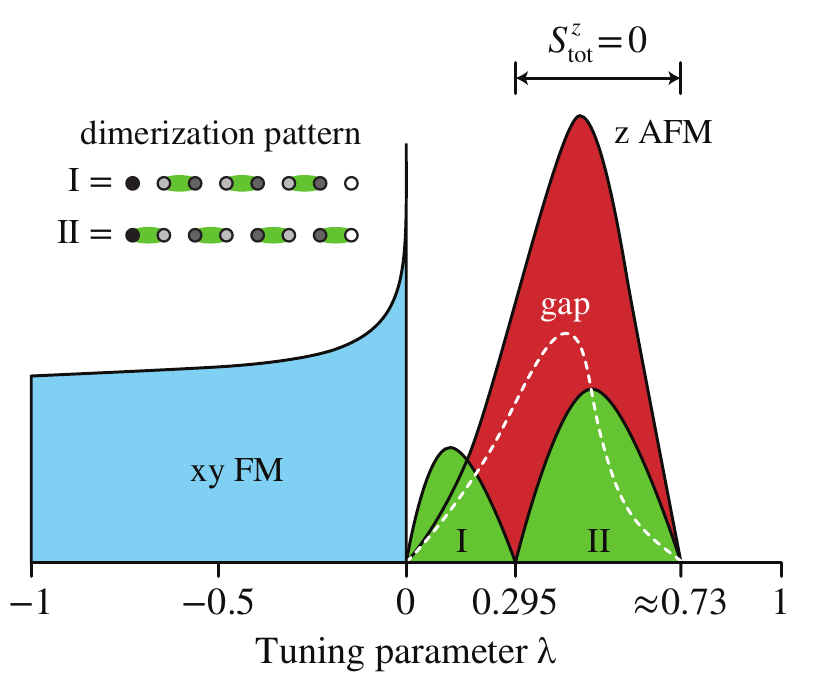}
\end{center}
\caption{\label{FIG:phase-diagram} This diagram summarizes the ground state 
properties of a spin-half chain with tunable interactions. The system is defined
on an open chain with oppositely directed, fully polarized boundary spins. The 
Fredkin model coincides with $\lambda = 0$; the extremal points describe a conventional
nearest-neighbor Heisenberg model with ferromagnetic ($\lambda = -1$) and antiferromagnetic couplings ($\lambda = 1$).
The xy ferromagnetism is enhanced as $\lambda$ increases from $-1$, reaching its maximum at $\lambda = 0$. But the Fredkin point serves as the phase boundary.
For $\lambda = 0^+$, the ferromagnetism is destroyed, giving way to a gapped, dimerized phase with z-directed Ising antiferromagnetic order. Two different dimer patterns are realized, which exclude (I) and include (II) the boundary spins.
The lowest-lying excitations are $S_{\text{tot}}^z = \pm 1$ in character, except in the parameter region coinciding with dimer pattern II, where they are $S_{\text{tot}}^z = 0$.}
\end{figure}

Systems with comparable character have now been created for 
odd-half-integer spin, at the cost of allowing for three-site interactions. 
Salberger and Korepin have proposed a 
model of $S=1/2$ spins on an open chain, in which the interactions
take the form of a singlet-pair projector that is correlated with the up or down character of the spin at 
a third, adjacent site~\cite{Salberger-arXiv-16}. They refer to their model 
as the ``Fredkin chain,'' since the action of the three-site operation is analogous 
to the logic of the Fredkin gate used in reversible computing. 

The Fredkin spin chain is frustration-free, and its ground state wave function is known exactly. Nonetheless, measuring most observables is nontrivial, and extrapolation to the 
thermodynamic limit requires careful accounting of the subleading corrections to finite-size scaling. In Sec.~\ref{SECT:Fredkin-Spin-Chain}, we present various schemes for evaluating spin correlation functions in the Fredkin ground state. We provide data from exact enumeration and from an efficient stochastic sampling scheme. We also present analytic results for the infinite chain, on the basis of which we deduce a closed-form expression for the z-directed spin profile.

Like the integer-spin Motzin model, the Fredkin model can be colored and promoted to higher spin values, but the original spin-half version has the virtue of obeying locality~\cite{DellAnna-PRB-16}. Other interesting generalizations and deformations of the model have been suggested
\cite{Zhang-JPA-17,Udagawa-JPA-17,Padmanabhan-arXiv-18}, including one from the present authors~\cite{Adhikari-APS-17}, which we discuss in Sec.~\ref{SECT:Generalized-Spin-Chain}. We take the point of view that the Fredkin model can be placed along a continuum of correlated-exchange models. 
We present ground-state measurements from exact diagonalization (ED) and density matrix renormalization group (DMRG) calculations.

In Sec.~\ref{SECT:Conclusions}, we conclude with a summary of our key findings. As shown in Fig.~\ref{FIG:phase-diagram}, we find evidence for several competing phases. We discover that an antiferromagnetic perturbation causes a dramatic collapse of the xy ferromagnetic correlations present in the Fredkin state. This transition is marked by the opening of an excitation gap and the appearance of coexisting dimer order and z-directed staggered magnetic order. An unusual feature of our model is that the equivalence classes~\cite{Salberger-arXiv-16,comment1} that characterize the Fredkin model re-emerge (in modified form) at a special anti-Fredkin point, deep in the antiferromagnetic side of the phase diagram.

\section{Fredkin Spin Chain}
\label{SECT:Fredkin-Spin-Chain}

\subsection{Model}

We consider a finite chain of $N$ spin-half objects.
In the chain's interior, the Hamiltonian 
\begin{equation} \label{EQ:bulk_hamiltonian}
H_\text{bulk} = \sum_{i=2}^{N-1} F_i
\end{equation}
is the sum of three-site operators
\begin{equation} \label{EQ:fredkin_operator}
F_i = U_{i-1}P_{i,i+1} +
P_{i-1,i}D_{i+1}.
\end{equation}
Here, $U_i = \tfrac{1}{2}(\mathbb{1}+\sigma^z_{i})$, $D_i = \tfrac{1}{2}(\mathbb{1}-\sigma^z_{i})$, and $P_{i,i+1} = \frac{1}{4}(\mathbb{1}-\vec{\sigma}_{i}\cdot\vec{\sigma}_{i+1})$ are lone-spin-up, lone-spin-down, and spin-singlet-pair projectors, 
with $\vec{\sigma} = (\sigma^x, \sigma^y, \sigma^z)$ denoting the Pauli matrices~\cite{comment2}.

\begin{figure}
\begin{center}
\includegraphics{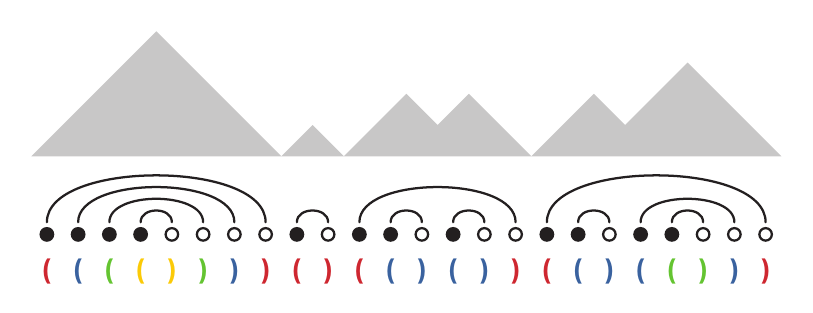}
\end{center}
\caption{\label{FIG:Fredkin} The ground state of the Fredkin spin chain is
an equal-weight superposition of spin states forming a balanced string.
This Dyck word condition can be viewed in various ways---as properly nested parentheses, 
as noncrossing bonds connecting up and down spins, or as a height profile
that never drops below the horizon.}
\end{figure}

Boundary terms, taking the form of magnetic fields applied independently 
to the leftmost ($i=1$) and rightmost ($i=N$) spins in the chain, act to select a unique, 
lowest-energy state from the otherwise highly degenerate ground state 
manifold. 
In particular, we suppose that the bulk Hamiltonian is augmented by $z$-directed fields $\alpha$ and $-\beta$, 
which are Zeeman-coupled to the edge spins:
\begin{equation}
\begin{split}
H_\text{boundary} &= \alpha D_1
+ \beta U_N\\
&= \frac{\alpha}{2}\bigl(\mathbb{1}-\sigma^z_1\bigr) + \frac{\beta}{2}\bigl(\mathbb{1}+\sigma^z_N\bigr).
\end{split}
\end{equation}
A z-directed ferromagnetic phase is realized when $\alpha$ and $\beta$ have opposite sign 
(i.e., when the fields are aligned). This is a trivial case with no 
spin twist along the length of the chain. More interesting phases arise
when $\alpha$ and $\beta$ have the same sign.
We focus on the case of $\alpha, \beta > 0$, which encourages a spin up (down) at the left (right) edge of the chain. The resulting ground state wave function 
is an equal-weight superposition 
\begin{equation} 
\ket{\psi_F} = \frac{1}{\sqrt{C_{N/2}}}\sum_\mathcal{D} \ket{\mathcal{D}}
\end{equation}
of spin configurations that form a balanced string. In these so-called Dyck word states, $\mathcal{D}$, the up and down spins are 
matched and perfectly nested. See Fig.~\ref{FIG:Fredkin}. 
For a given chain length $N$, the total number of such states
is $C_{N/2} = \frac{1}{N/2+1}{N \choose N/2}$, where $C_n$ is the Catalan number.

At very small system sizes, the ground state mostly has the character of a
domain wall state. As $N$ increases, however, the spins away from the boundaries flop into the xy plane. In the thermodynamic limit, the ferromagnetic correlations deep in the bulk are at maximum strength, i.e.,
$\expectation{\sigma^x_i \sigma^x_j + \sigma^y_i \sigma^y_j} = 1$,
where $i,j$ label spins away from the chain edges. [See the arguments preceding Eq.~(29) in Ref.~\onlinecite{Chen-JPA-17}.]

While the nature of the ground state depends only on the sign of $\alpha$ and $\beta$, the spectrum of excited
states depends on their specific values. For concreteness, we consider the extreme
limit in which $\alpha$ and $\beta$ are both taken to be arbitrarily large. In practice, this amounts to working in a restricted Hilbert space such that the spin at the left edge of the chain is fixed in the state $\ket{\uparrow_1}$ and the spin at the right edge is $\ket{\downarrow_N}$. 
The terms in $F_2$, for example, simplify to 
$U_1 P_{2,3} = P_{2,3}$ and
\begin{equation}
\begin{split}
P_{1,2} D_3 &= \frac{1}{4}(\mathbb{1}-\vec{\sigma}_1\cdot\vec{\sigma}_2) D_3 \\
&= \frac{1}{4}(\mathbb{1}-\sigma^z_2) D_3
= \frac{1}{2}D_2D_3.
\end{split}
\end{equation}
Hence, the effective Hamiltonian in this limit is
\begin{equation}
H = \sum_{i=3}^{N-2}F_i + P_{2,3} + \frac{D_2D_3}{2} + P_{N-2,N-1} + \frac{U_{N-2}U_{N-1}}{2}.\end{equation}
This formalizes the statement that our model for the length-$N$
chain consists of $N-2$ active interior spins and two frozen, fully polarized edge spins. 
In all our numerical work, we have verified that the results we obtain
for the ground state and few lowest-lying states
are consistent with setting finite field values $\alpha = \beta = 1$
(which is the choice made in Ref.~\onlinecite{Salberger-arXiv-16}).

\subsection{Measuring observables in the ground state}

The basic technical challenge for evaluating the properties of the Fredkin model ground state is to generate its $C_{N/2}$ spin configurations of Dyck word form---either exhaustively or via sampling. As it turns out, the Dyck words have a natural lexical ordering, and there are well developed algorithms for systematically stepping through all words of a given length~\cite{Knuth-Addison-97,Jorg-Springer-11}. This is all one needs to make any diagonal or off-diagonal spin measurements: e.g., 
\begin{equation} 
\bra{\psi_F} \sigma^z_i \ket{\psi_F} = \frac{1}{C_{N/2}}
\sum_{\mathcal{D}} \bra{\mathcal{D}} \sigma^z_i \ket{\mathcal{D}}
\end{equation}
and
\begin{equation}
\begin{split}\bra{\psi_F} \bigl( & \sigma^+_i \sigma^-_j + \sigma^-_i \sigma^+_j\bigr) \ket{\psi_F}\\
 &= \frac{1}{C_{N/2}}
\sum_{\mathcal{D},\mathcal{D}'} \bra{\mathcal{D}'} \bigl(\sigma^+_i \sigma^-_j + \sigma^-_i \sigma^+_j\bigr) \ket{\mathcal{D}} \\
&= \frac{1}{C_{N/2}}
\sum_{\mathcal{D}}\begin{cases}
1 & \text{if $\downarrow_i$ and $\uparrow_j$,}\\
1 & \text{if $\uparrow_i$ and $\downarrow_j$ and Dyck word form}\\
& \text{is preserved after spins are flipped,}\\
0 & \text{otherwise,}
\end{cases}
\end{split}
\end{equation}
with $2\sigma^{\pm} = \sigma^x \pm i \sigma^y$ defining
the raising and lowering operators.
Nonetheless, exact enumeration of systems much larger than $N=38$ is impractical because of memory constraints.

A good alternative is to consider a sampling algorithm that generates Dyck words stochastically. Here, we propose such an algorithm. We proceed by defining a height function
\begin{equation} 
h_i = \sum_{j=1}^i \sigma^z_j. 
\end{equation}
The set of Dyck word states corresponds to all specifications of spins $\{ \sigma^z_1, \sigma^z_2, \ldots, \sigma^z_N \}$ such that $h_0 = h_N = 0$ and $h_i \ge 0$ for all $i$. In other words, the height function defines a landscape that never drops below the horizon.

We then make the analogy with a biased random walk ($h_{i+1} = h_i \pm 1$) restricted to one-sided excursions ($h_i \ge 0$) and attempt to generate each valid landscape recursively from left to right. 
The following branching probabilities ensure that each walk starting from $h_0=0$ never drops below the horizon and returns to height zero
after exactly $N$ steps:
\begin{equation} \begin{split}
\text{Prob}(\sigma^z_{i+1} = +1) &= \frac{(h_i+2)(N-i-h_i)}{2(h_i+1)(N-i)},\\
\text{Prob}(\sigma^z_{i+1} = -1) &= \frac{h_i(N-i+h_i+2)}{2(h_i+1)(N-i)}.
\end{split} \end{equation}
Note that taking the limit $N \to \infty$ in these formulas correctly accounts for the infinite-chain limit. Hence we are able to use Monte Carlo techniques to treat finite-size systems or to simulate the thermodynamic limit directly.

\begin{figure}
\includegraphics[width = 0.85\columnwidth]{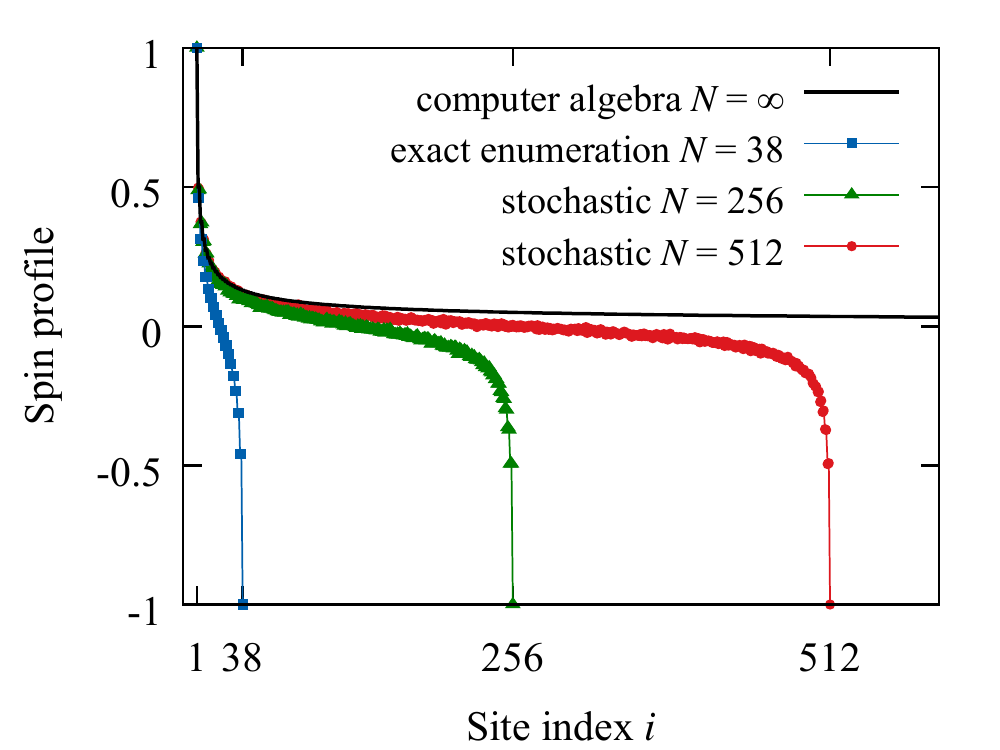}
\includegraphics[width = 0.85\columnwidth]{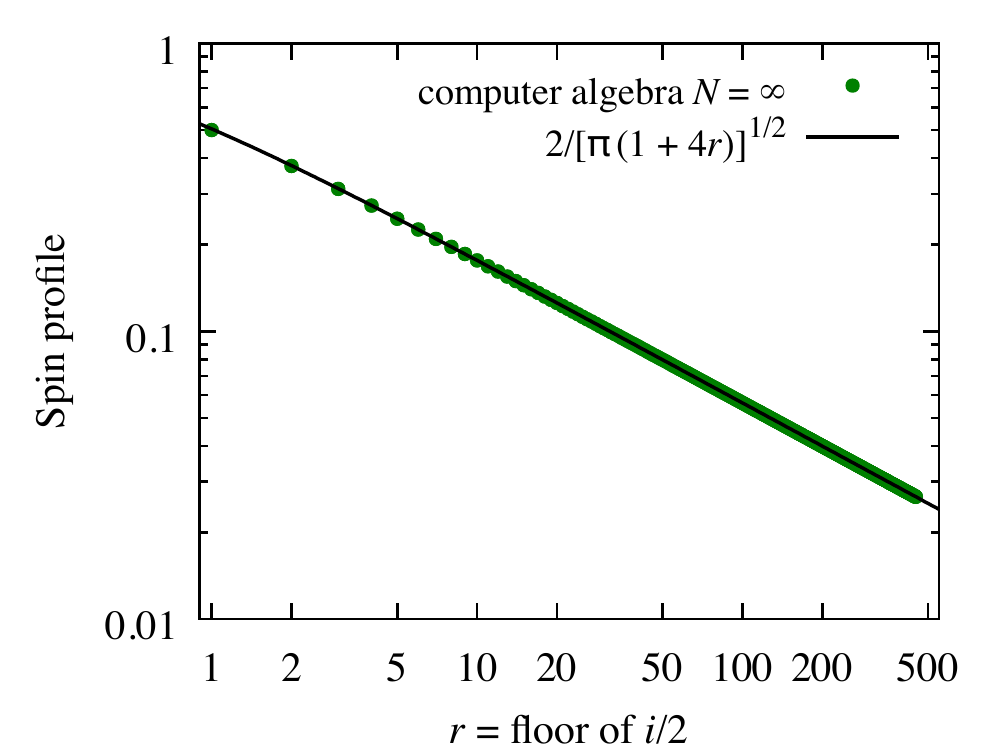}
\caption{\label{FIG:fredkin_spin_profile}Top: The z-directed spin profile is shown for various finite system sizes and for the infinite-chain limit.
Bottom: Fits of the asymptotic behavior suggest that the spin profile
falls off as one over the square root of distance from the edge spin
in the infinite-chain limit.
}
\end{figure}

The random walk analogy is useful in one other way. It serves as a book-keeping trick for tallying all spin configurations recursively from the left edge rightward. We find that 
any diagonal spin measurement consisting of a product of $\sigma^z_i$ taken over some finite range can be solved analytically. For instance, using computer algebra, we find that
\begin{equation} \label{EQ:spin-profile-fractions}
\begin{split}
\expectation{\sigma^z_1} &= 1\\
\expectation{\sigma^z_2} = \expectation{\sigma^z_3} &= \frac{N-4}{2(N-1)} \to \frac{1}{2}\\
\expectation{\sigma^z_4} = \expectation{\sigma^z_5} &= \frac{3(N^2-10N+16)}{8(N-3)(N-1)} \to \frac{3}{8}\\
\expectation{\sigma^z_6} = \expectation{\sigma^z_7} &= \frac{5(N^3-18N^2+80N-96)}{16(N-5)(N-3)(N-1)} \to \frac{5}{16}
\end{split}
\end{equation}
so that each value $\expectation{\sigma^z_i}$ is given by a ratio of polynomials of order $\lfloor i/2 \rfloor$ in $N$; the limiting values shown in Eq.~\eqref{EQ:spin-profile-fractions} correspond to $N\to \infty$. We have carried out such computations for a range of site indices $i=1,2,\ldots, 900$. The results lead us to conclude that the spin profile in the thermodynamic limit (shown in the bottom panel of Fig.~\ref{FIG:fredkin_spin_profile}) has the following closed-form expression:
\begin{equation}
\expectation{\sigma^z_i} = \frac{1}{2^{2r}}{ 2r \choose r} = \frac{2}{\sqrt{\pi(1 +  4r)}} + O(r^{-5/2}),
\end{equation}
where $r = \lfloor i/2 \rfloor$. The asymptotic form $1/\sqrt{\pi r}$ has previously been derived from an effective theory of the continuum height field; see Eq. (23) of Ref.~\onlinecite{Chen-JPA-17}.

\begin{figure}
\includegraphics[width = 0.85\columnwidth]{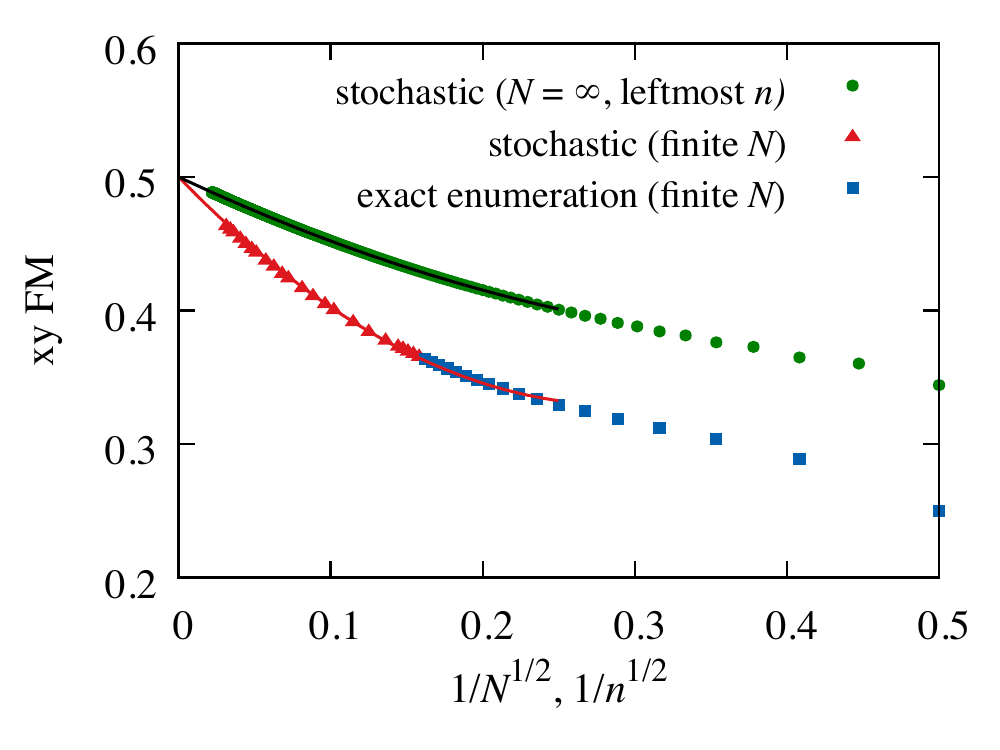}
\caption{\label{FIG:fredkin_fm}
The order parameter for ferromagnetic alignment in the xy plane, $\sum_{i,j}\expectation{\sigma_i^+\sigma_j^-
+ \sigma_i^-\sigma_j^+}_N$, is plotted against the inverse square-root system size. The solid red line is a second-order polynomial through the data $(32 < N \le 1024)$. The solid black line is a fit of the same kind, but taken through data for the cumulative xy ferromagnetism,
$\sum_{i,j \le n}\expectation{\sigma_i^+\sigma_j^-
+ \sigma_i^-\sigma_j^+}_{\infty}$,
 measured from one end of the infinite chain. Both extrapolations agree in suggesting that the xy ferromagnetic correlations in the interior of the Fredkin chain are saturated.
That is to say, the limits $\expectation{\sigma^z_i} \to 0$ and 
$\expectation{\sigma^x_i \sigma^x_j + \sigma^y_i \sigma^y_j} \to 1$ are achieved rapidly as a function of distance from the chain edges.}
\end{figure}

We now emphasize an important point regarding the magnetic properties in the ground state. Udagawa and Katsura~\cite{Udagawa-JPA-17},
following the reasoning in Ref.~\onlinecite{Movassagh-JMP-17},
argue that the ``magnetization in the ground state is along the z-direction'' because $\expectation{\sigma^x_i} = \expectation{\sigma^y_i} = 0$. This is somewhat misleading. While it is true that expectation values of $\sigma^x_i$ and $\sigma^y_i$ vanish (trivially, by symmetry) in the ground state for any finite-length chain, the more relevant issues are whether there are long-range spin correlations in the ground state (yes, strongly ferromagnetic in the easy xy plane) and whether symmetry-breaking order appears in the thermodynamic limit (no, by virtue of the Mermin-Wagner-Hohenberg theorem~\cite{Mermin-PRL-66,Hohenberg-PR-67}).

More specifically, the operator
\begin{equation}
\begin{split} 
m_\perp &= \sum_i \bigl[(\cos\theta)\sigma^x_i + (\sin\theta)\sigma^y_i\bigr] \\
&= \sum_i(e^{-i\theta}\sigma^+_i + e^{i\theta}\sigma^-_i),
\end{split}
\end{equation}
which measures the xy ferromagnetism,
satisfies $\expectation{m_\perp} = 0$, but we can
show that 
$\expectation{m_\perp^2} = \sum_{i,j}\expectation{\sigma_i^+\sigma_j^-
+ \sigma_i^-\sigma_j^+}$
saturates to a finite, nonzero value as $N\to\infty$.
Alternatively---rather than compute the ferromagnetic order parameter over an entire finite-length-$N$
chain---we can compute it over the leftmost $n$ sites of the infinite chain
$\sum_{i,j \le n}\expectation{\sigma_i^+\sigma_j^-
+ \sigma_i^-\sigma_j^+}_{\infty}$.
Both approaches give extrapolated values that are consistent with fully saturated xy spin correlations in the bulk (see Fig.~\ref{FIG:fredkin_fm}).

\begin{figure}
\begin{center}
\includegraphics[width = 0.85\columnwidth]{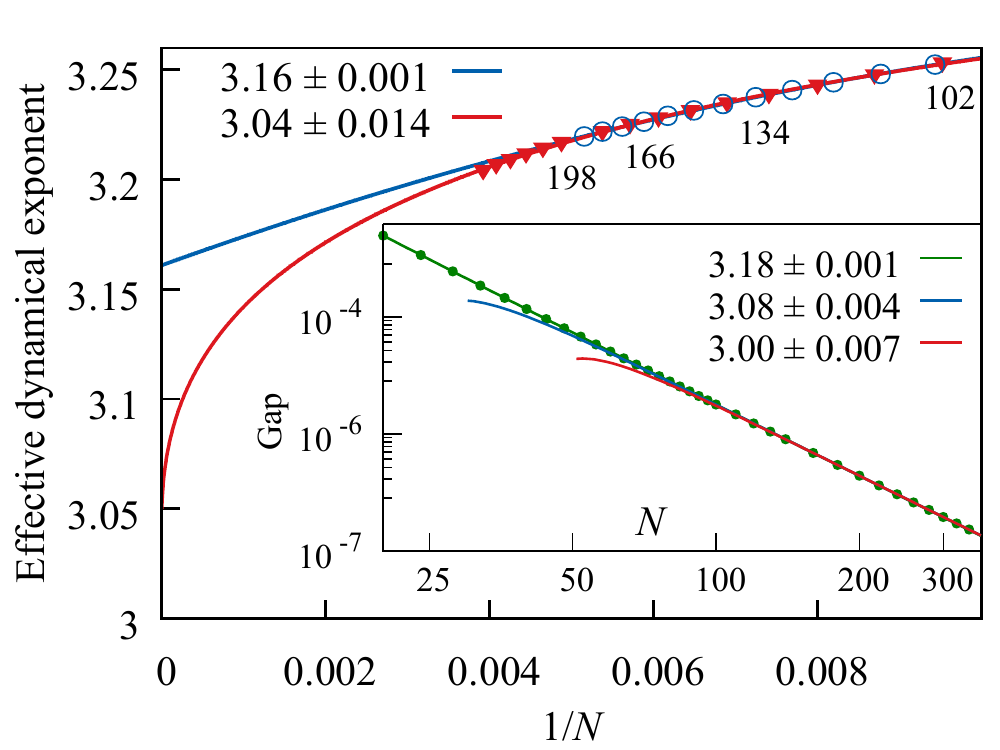}
\end{center}
\caption{\label{FIG:zExp} 
These data come from density matrix renormalization group 
calculations, which are described in Sec.~\ref{SUBSECT:Numerical-methods}.
In the main panel, the effective dynamical exponent for system size $N+\delta n$, determined from the measured gaps at sizes $N$ and $N+2\delta n$, is extrapolated to the thermodynamic limit.
The first extrapolation (blue, upper curve) is quadratic in $1/N$ (of the form $z_\infty + a_1/N + a_2/N^2$)
and fit to data (blue, open circles) on system sizes up to 200 with a finite difference $\delta n = 4$; this is comparable to the analysis carried out in Ref.~\onlinecite{Chen-JPA-17}.
The second (red, lower curve) allows for a powerlaw form and logarithmic corrections
[of the form $z_\infty + (b_0 + b_1\ln N)/N^\alpha$]
and is based on data up to size 260 (red, filled triangles) in steps of 10.
The inset shows three fits of the form
$\Delta(N) = N^{-z-c_1/N}\bigl(d_0 + d_1/N\bigr)$,
one running through the gap data for all measured system sizes
and the other two fit with lower size cutoffs of 150 and 200.
}
\end{figure}

This state of affairs is possible because the system supports gapless excitations. These are spin-wave-like in the xy spin plane but have the form of a back-and-forth sloshing of the domain wall when viewed in the z direction.
Exactly how the gap vanishes is of special interest. 
The energy difference between the lowest-lying ($S^\text{tot}_z = \pm 1$) excited states and the ($S^\text{tot}_z = 0$) ground state 
scales as $\Delta \sim N^{-z}$, with a dynamical exponent that is unusually
large. (This is essentially a jamming effect; the Fredkin term $-F_j$ is a short-bond-shuffle operation, so the evolution is highly constrained.) Movassagh has derived~\cite{Movassagh-AMSA-18} strict upper and lowers bounds, $2 < z < 15/2$. 
The current best numerical estimate, $z=3.2$~\cite{Chen-JPA-17}, lies toward the lower end of this range. 

Our own numerics are not inconsistent with 3.2, but we stress that a wide range of exponents between 2.8 and 3.3 can be achieved with plausible fitting functions, depending on the precise form of the subleading corrections assumed and the size cutoffs that controls which data are included in the fit. Figure~\ref{FIG:zExp} showcases various efforts to extract the dynamical exponent. We consider both (i)
$z_\text{eff}(N+2) = [\ln \Delta(N+4)/\Delta(N)]/[\ln N/(N+4)]$,
extrapolated in various ways to $1/N=0$, 
and (ii) a single fit
of the full gap data set to the function
$\Delta(N) = N^{-z-c_1/N}\bigl(d_0 + d_1/N\bigr)$.
At most, we can say that the correct value of the dynamical
exponent lies in the range $3.0 \lesssim z \lesssim 3.2$.

\section{Generalized Spin Chain}
\label{SECT:Generalized-Spin-Chain}

\subsection{Model}

The conventional Heisenberg model involves two-body interactions between neighboring spins, and those interactions are SU(2) invariant. The Fredkin model, on the other hand, makes use of three-body terms that break the spin-rotation-symmetry by picking out a special direction in spin space.
We have proposed a tunable model~\cite{Adhikari-APS-17,Adhikari-APS-18} (in the parameter $\lambda$) that interpolates between the Fredkin spin chain (at $\lambda = 0$) and the conventional ferromagnetic ($\lambda = -1$) and antiferromagnetic ($\lambda = 1$) quantum Heisenberg models. Recent work has appeared~\cite{Chen-JPA-17} that addresses the ferromagnetic side ($\lambda < 0$) of this model.

The essence of the generalization is to apply to 
Eq.~\eqref{EQ:fredkin_operator} the operator replacement
\begin{equation} \label{EQ:operator_replacement}
\begin{split}
U_i &\to L_i(\lambda) = \biggl(\cos\frac{\pi\lambda}{2}\biggr)U_i -\biggl(\sin\frac{\pi\lambda}{2}\biggr) \mathbb{1},
\\
D_i &\to R_i(\lambda) = \biggl(\cos\frac{\pi\lambda}{2}\biggr)D_i -\biggl(\sin\frac{\pi\lambda}{2}\biggr) \mathbb{1}.
\end{split}
\end{equation}
Hence, the basic three-site operation in the interior is
\begin{equation} \label{EQ:Generalized_Model}
G_i(\lambda) =L_{i-1}(\lambda) P_{i,i+1}  \\
+P_{i-1,i} R_{i+1}(\lambda).
\end{equation}
For $\lambda = 0$, we recover the Fredkin chain, since $L_i(0) = U_i$ and $R_i(0) = D_i$ and hence $G_i{(0)} =  F_i$.
Note that any deviation away from this special tuning point breaks the purely Dyck word structure of the ground state, and for $\lambda > 0$ the model is no longer frustration-free.

For $\lambda = \mp 1$, the operators $L_i(\mp 1) = R_i(\mp 1) = \pm \mathbb{1}$ are proportional to the unit matrix, so the model reverts to a ferromagnetic or antiferromagnetic Heisenberg model with  conventional two-body exchange interactions:
\begin{equation}
\label{EQ:Gm_pm_1}
\sum_i G_i{(\mp1)} =  \pm\sum_i \bigl(P_{i,i+1}+P_{i-1,i}\bigr) \simeq \pm 2\sum_iP_{i,i+1}.
\end{equation} 

The operators $L_i(\lambda)$ and $R_i(\lambda)$, defined as linear combinations in Eq.~\eqref{EQ:operator_replacement}, act as dynamical exchange couplings with different magnitude and sign depending on the spin-up or spin-down character of the measured spin at site $i$.
As $\lambda$ increases from $0$, the $L_i(\lambda)_{\uparrow\uparrow}$ entry decreases from 1, and $L_i(\lambda)_{\downarrow\downarrow}$ becomes nonzero and increasingly negative.
At $\lambda_\text{c1} = 0.295$, where $2\tan(\pi\lambda_\text{c1}/2) = 1$, the spin-up and spin-down entries act with equal magnitude but opposite sign:
\begin{equation}
L_i{(\lambda_\text{c1})} = \frac{1}{\sqrt{5}}\begin{pmatrix} 1 & 0 \\ 0 & -1 \end{pmatrix} = \frac{1}{\sqrt{5}}[U_i-D_i] .\\
\end{equation}
As $\lambda$ is tuned higher still, $L_i(\lambda)_{\uparrow\uparrow}$ continues to decrease. At $\lambda = 0.5$, it vanishes entirely:
\begin{equation}
L_i(0.5) =  \frac{1}{\sqrt{2}}\begin{pmatrix} 0 & 0 \\ 0 & -1 \end{pmatrix} = -\frac{1}{\sqrt{2}}D_i.
\end{equation}
At this parameter value, which we call the anti-Fredkin point, the resulting three-site interaction is similar to that at the Fredkin point, but the sign of the exchange and the filtering at the third adjacent site is reversed. That is to say, whereas the Fredkin model applies a ferromagnetic exchange that acts between a pair of sites whenever a spin up is to its left or a spin down to its right, the anti-Fredkin interaction is antiferromagnetic and acts when there is a spin down to the left and spin up to the right.
Equation~\eqref{EQ:Generalized_Model} can also be viewed as a linear superposition of the Fredkin model $G_i{(0)}$ and the Heisenberg models $G_i{(\mp1)}$:
\begin{equation}
G_i(\lambda) = \biggl(\cos\frac{\pi\lambda}{2}\biggr)G_i{(0)} \mp \biggl(\sin\frac{\pi\lambda}{2}\biggr)G_i{(\mp 1)}.
\end{equation}

\begin{figure}
\begin{center}
\includegraphics[width = 0.85\columnwidth]{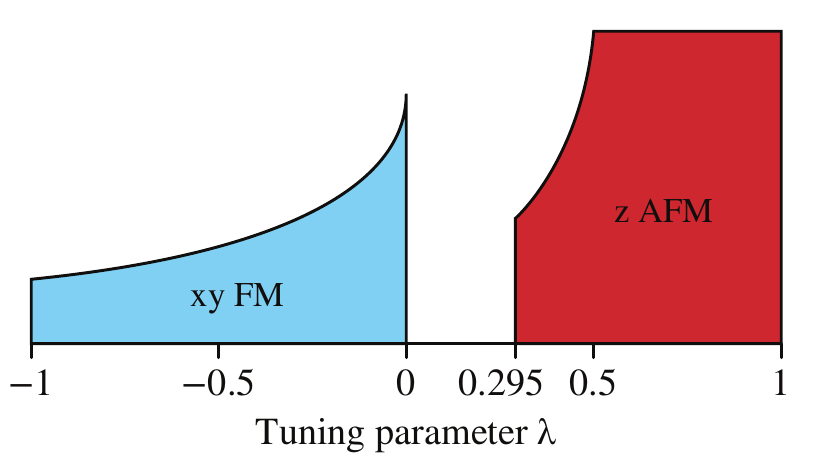}
\end{center}
\caption{\label{FIG:classical-phase-diagram} The phase diagram of the corresponding classical model is drawn based on the measurements of the ground state spin configuration obtained from a global energy optimization. The optimized spin texture shows no direct connection between FM and AFM phases,
but instead presents an intervening sequence of
spin textures (reminiscent of micro-emulsions~\cite{Spivak-PRB-04,Jamei-PRL-05})
with highly idiosyncratic patterns whose
repeat unit grows in steps from 1 to the full system size.
}
\end{figure}

We note that in the classical version of this model---with $\vec{\sigma}$ passing over to a continuous unit vector---the ground state is a fully saturated, z-directed antiferromagnet (z AFM) for $1/2 < \lambda \le 1$; the antiferromagnetic order is canted when $\lambda_\text{c1} < \lambda \le 1/2$; and there is strong, xy-directed ferromagnetism (xy FM) for $\lambda \leq 0$. The phase diagram is depicted in Fig.~\ref{FIG:classical-phase-diagram}. In Sects.~\ref{SUBSECT:Numerical-methods} and \ref{SUBSECT:Results}, we will show that the quantum version, with spin-half degrees of freedom, looks similar on the ferromagnetic side ($\lambda \le 0$)---albeit disordered by U(1) Goldstone modes---but supports a quite different collection of ordered phases on the antiferromagnetic side ($\lambda > 0$).

\subsection{Numerical methods}
\label{SUBSECT:Numerical-methods}

For a collection of $N$ spin-half objects, the dimension of the Hilbert space is $(2S+1)^N = 2^N$. Because of this exponential scaling with system size, ED studies are limited to rather small lattices, a few tens of spins at most. This is true even if we take advantage of all the available symmetries and use a Lanczos algorithm to compute just a few of the lowest-lying states. For the model under consideration, we are hindered by the fact that there are only a small number of symmetries that can be exploited.

 The Hamiltonian commutes with $S^z_\text{tot}
= (1/2)\sum_{i=1}^N \sigma^z_i$, the z-component of
the total spin, but it does not commute with $S_\text{tot}^2
= \vec{S}_\text{tot} \cdot \vec{S}_\text{tot} = (1/4)\sum_{i,j} \vec{\sigma}_i\cdot\vec{\sigma}_j$, the total spin magnitude. 
The model possesses a symmetry in which the lattice is inverted 
and all spins are reversed; in other words, the model is invariant under the transformation
\begin{equation}
\sigma^a_i \Leftrightarrow \bigl(\sigma^a_{L+1-i}\bigr)^T,
\end{equation}
for each of $a=x,y,z$. Right at the Fedkin point, we can take advantage of the equivalence classes defined by the number of Dyck-word mismatches, but that does not apply away from $\lambda=0$. Finally, because the model is defined on an open chain, there is no translational symmetry and hence no well-defined crystal momentum quantum number. So, at best, the Hamiltonian matrix can be block-diagonalized in $S_{\text{tot}}^z$ and one additional Z$_2$ quantum number. This basis-size reduction is not enough to significantly reduce the computational cost. Still, to help guide our investigations, we have generated the full set of energy eigenstates for $N = 12, 14, 16, 18$ over a densely spaced mesh of $\lambda$ values.

To access larger system sizes, we make use of a DMRG algorithm implemented in the open-source C++ library ITensor~\cite{ITensor}. We are mindful of the fact that the high level of entanglement in the vicinity of $\lambda = 0$ requires some additional care. Accordingly, we have employed a very conservative convergence criterion: the DMRG algorithm runs through $7N+30$ sweeps using an adaptive truncation cutoff at relative error $10^{-12}$ with maximum bond dimension $10 N$. We have verified our DMRG results to near double-precision floating-point accuracy against ED results for $N \leq 18$.

The DMRG computation is carried out for all even lattice sizes up to $N=80$, over a tight mesh of tuning parameter values $\lambda=-1.000,-0.995,-0.990, \ldots, 0.995, 1.000$. The most expensive of those simulations corresponds to 590 sweeps with a maximum bond dimension of 800. The one exception is at $\lambda=0$, the Fredkin point, where we have made an extra effort to simulate system sizes in steps of 10 up to $N=350$. Here, the most expensive simulation corresponds to 2520 sweeps with a maximum bond dimension of 3500. Various physical quantities are computed in the ground state as a function of the tuning parameter $\lambda$; i.e., $O(\lambda) = \expectation{\hat{O}} = \bra{\psi_0(\lambda) }\hat{O}\ket{\psi_0(\lambda)}$. These include the spin profile $\expectation{\sigma_i^z}$, dimer profile $\expectation{\sigma_i^z \sigma_{i+1}^z}$, dimer order parameter
\begin{equation}
\expectation{d_{\parallel}^2} = \frac{1}{N^2}\sum_{i,j=1}^{N-1}(-1)^{i+j}\expectation{\sigma_i^z \sigma_{i+1}^z\sigma_j^z \sigma_{j+1}^z},
\end{equation}
Ising antiferromagnetic order parameter (z AFM) 
\begin{equation}
\expectation{m_{\parallel}^2} = \frac{1}{N^2}\sum_{i,j=1}^N (-1)^{i+j}\expectation{\sigma_i^z\sigma_j^z}
\end{equation}
(as well as its Binder cumulant $Q = 1-{\expectation{m_{\parallel}^4}}/{3\expectation{m_{\parallel}^2}^2}$), and the xy-plane ferromagnetic order parameter (xy FM) 
\begin{equation}
\expectation{m_{\perp}^2} = \frac{1}{N^2}\sum_{i,j=1}^N \expectation{\sigma_i^+\sigma_j^-+\sigma_i^-\sigma_j^+}.
\end{equation}

\subsection{Results}
\label{SUBSECT:Results}

\begin{figure}
\begin{center}
\includegraphics[width = 0.85\columnwidth]{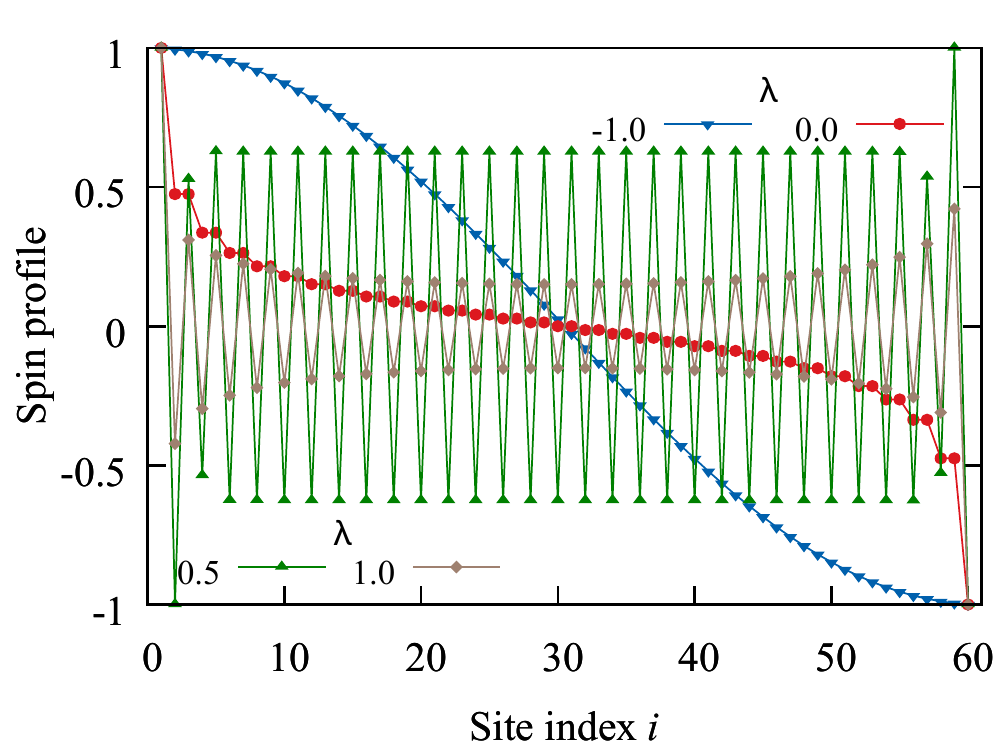}
\end{center}
\caption{\label{FIG:SpinProfile}The spin profile $\expectation{\sigma^z_i}$ gives the average projection of the spins along the z axis, spatially resolved over sites $i=1,2,\ldots, N$. Results are given here for system size $N=60$ and parameter values $\lambda = -1$, $0$, $0.5$, and $1$. Expectation values are computed in the ground state wave function, as determined by DMRG.
}
\end{figure}

At the Fredkin point, the ground state is an equal-amplitude superposition of all spin configurations of Dyck word form. Away from $\lambda = 0$, other non-Dyck-word spin configurations contribute; the Dyck word states still predominate, but they appear with idiosyncratic, $\lambda$-dependent amplitudes. Everywhere on the ferromagnetic side of the phase diagram ($\lambda \le 0$), the single configuration with the largest weight is (((($\cdots$)))); on the antiferromagnetic side ($\lambda > 0$), it is ()()$\cdots$()().
At two special points, $\lambda_\text{c1} = 0.295$ and $\lambda = 0.5$, the hard-domain-wall configuration (((($\cdots$)))) has no contribution.
The change in relative contributions manifests itself in the spin profile, which is shown in Fig.~\ref{FIG:SpinProfile}.

\begin{figure*}
\begin{center}
\includegraphics[width = 0.925\textwidth]{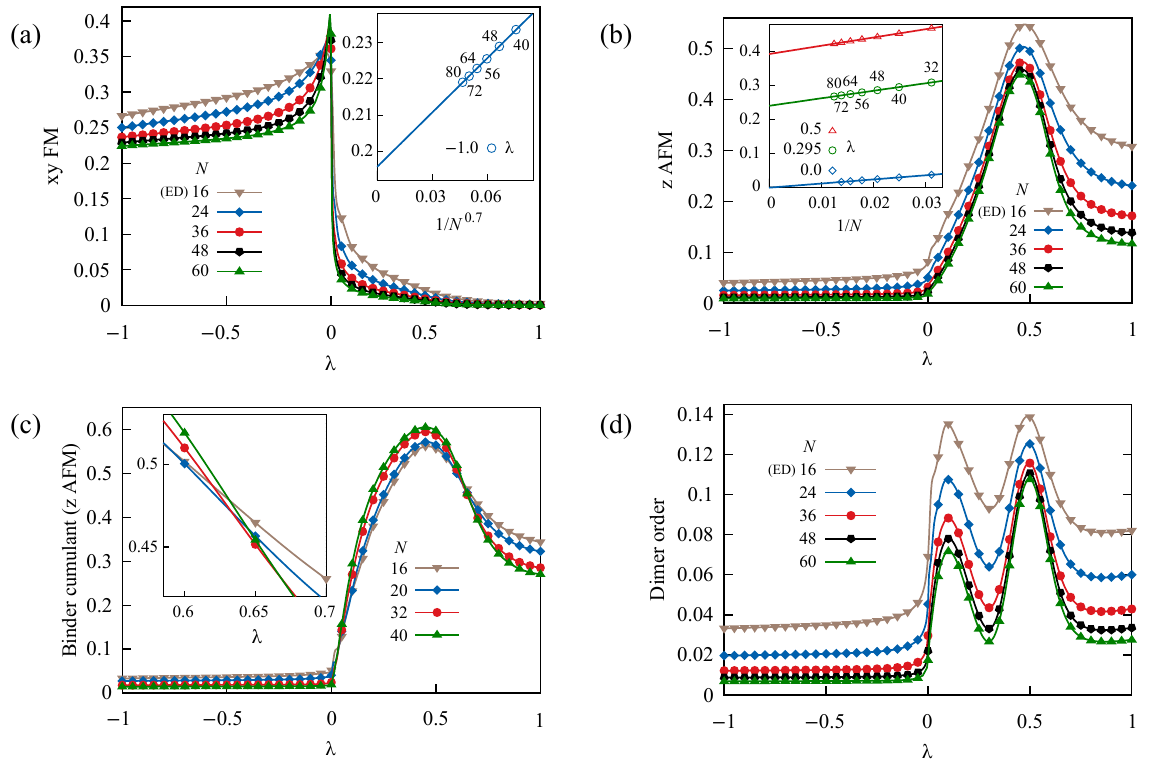}
\end{center}
\caption{\label{FIG:merged_order_params}(a) The xy-directed ferromagnetism, from DMRG and ED calculations, is plotted as a function of $\lambda$ for various system sizes $N$. The inset plot in the top right shows the data for $\lambda = -1$ extrapolated versus $1/N^{0.7}$ to $0.196(3)$ in the thermodynamic limit. The numeric labels in the inset plot denote selected values of the system size.
(b) The z-directed antiferromagnetism is plotted versus $\lambda$ for various system sizes $N$. The maximum value $0.4138(1)$ is obtained at $\lambda = 0.455(2)$. The inset shows extrapolations in $1/N$ for the values $\lambda = 0$, $0.295$, and $0.5$. The system has magnetic order in z-direction for $\lambda = 0^+$ in the thermodynamic limit. (c) The Binder cumulant for the Ising antiferromagnetism is plotted as a function of $\lambda$ for system sizes $N = 16$, $20$, $32$, and $40$. The crossing points mark the phase boundaries of the region with z-directed staggered order. The inset illustrates the drift of the rightmost crossing point because of strong subleading corrections to finite-size scaling. (d) The dimer order is plotted as a function of $\lambda$ for various system sizes $N$. The two lobes represent two phases with distinct patterns of dimerization.
}
\end{figure*}

Because of the incompatible boundary conditions, a ferromagnetic state with full spin-rotation symmetry cannot form at $\lambda = -1$.
Instead, the system undergoes a spin flop into the xy plane.
The xy-directed ferromagnetism persists all the way up to $\lambda = 0$. Along the way, the three-body interaction term conspires to produce a strong enhancement of the spin alignment [see Fig.~\ref{FIG:merged_order_params}(a)]. 

In the thermodynamic limit, the long-range ferromagnetic correlations die immediately as $\lambda \to 0^+$, making way for a new phase in which dimer order and z-directed, staggered magnetic order coexist. 
The numerical data to support this claim are shown in Figs.~\ref{FIG:merged_order_params}(a)--(d).
The $\lambda > 0$ side of the phase diagram exhibits two 
different patterns of dimerization, which are distinguished by a one-site shift relative to the boundary spins. At large values of the tuning parameter, $0.73 \lesssim \lambda \le 1$, all order vanishes. This region of the phase diagram is a disordered spin liquid, continuously connected to the ground state of the conventional nearest-neighbor, antiferromagnetic quantum Heisenberg model.

\begin{figure}
\begin{center}
\includegraphics[width = 0.85\columnwidth]{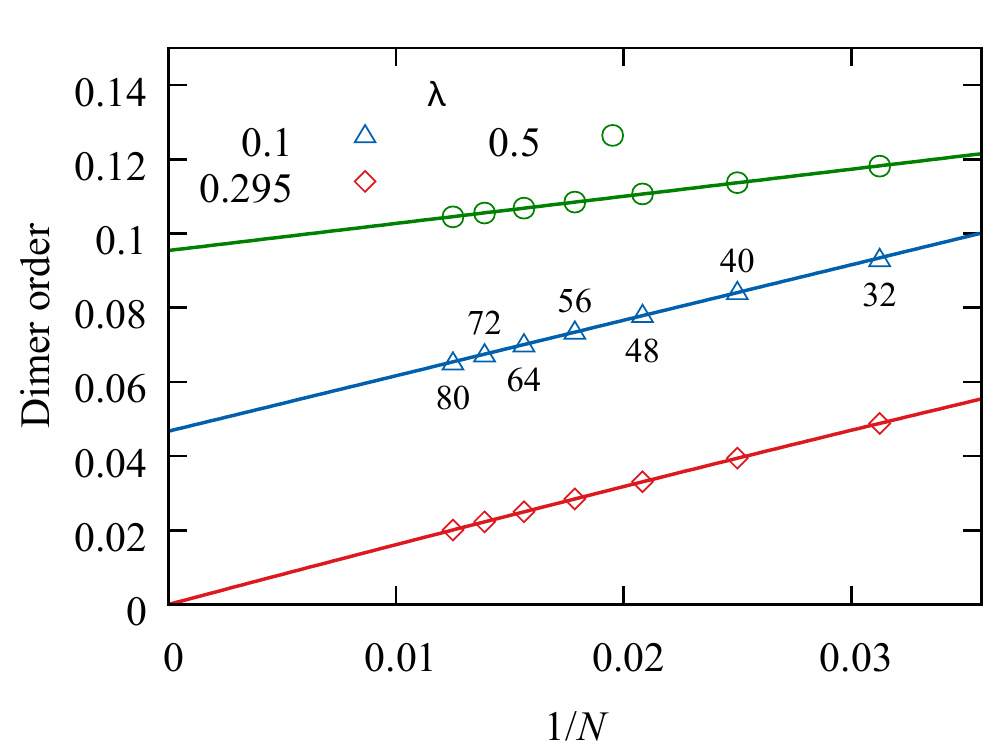}
\end{center}
\caption{\label{FIG:DimerFit}The dimer order in the thermodynamic limit is determined for parameter values $\lambda = 0.1$, $0.295$, and $0.5$. The values $\lambda = 0.1$ and $\lambda = 0.5$ are chosen to coincide with the two maxima visible in Fig.~\ref{FIG:merged_order_params}(d). The extrapolated values are $0.0468(2)$ at $\lambda = 0.1$ and $0.0954(1)$ at $\lambda = 0.5$. The dimer order appears to vanish continuously at $\lambda_\text{c1} = 0.295$.}
\end{figure}

\begin{figure}
\begin{center}
\includegraphics[width = 0.85\columnwidth]{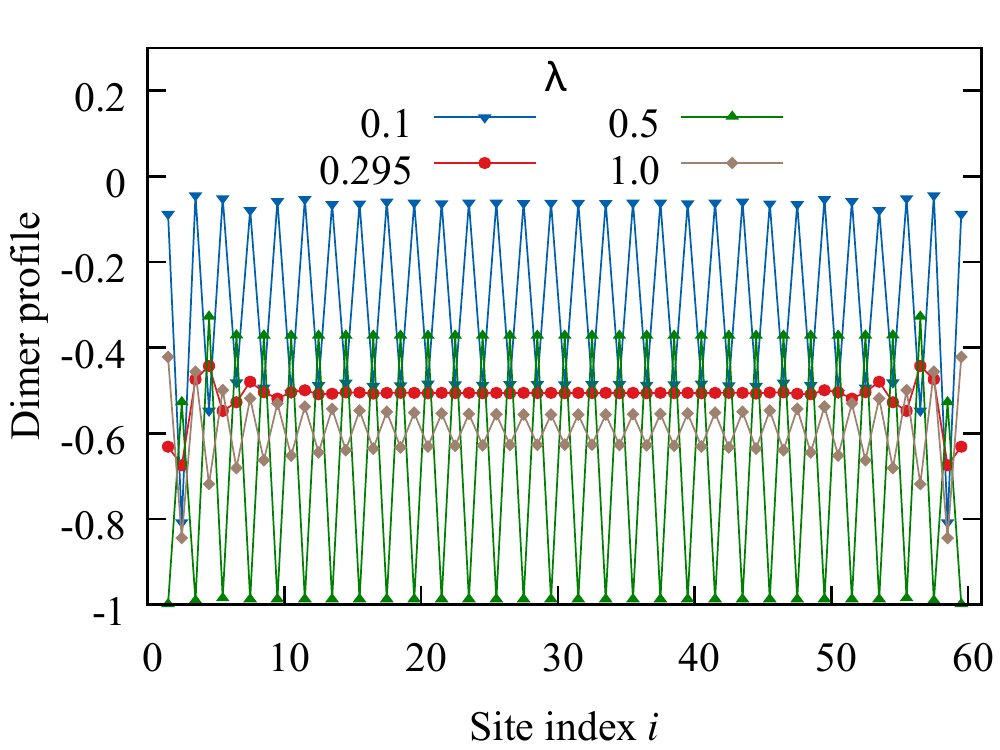}
\end{center}
\caption{\label{FIG:DimerProfile}The dimer profile is the two-point correlation function $\expectation{\sigma_i^z\sigma_{i+1}^z}$, measured along the spin chain. Results are given here for system size $N=60$ and parameter values $\lambda = 0.1$, $0.295$, $0.5$, and $1$. Expectation values are computed with respect to the ground state wave function, as determined by DMRG. The dimer profile reveals two different pattern of dimerization. The magnitude of the correlation function at its peak value (at $\lambda = 0.5$) is fully saturated (value $-1$) on alternating bonds.
}
\end{figure}

Figure~\ref{FIG:merged_order_params}(d) reveals the two lobes of dimer order (Fig.~\ref{FIG:DimerFit}), corresponding to the patterns plotted in Fig.~\ref{FIG:DimerProfile} and illustrated in the top left of Fig.~\ref{FIG:phase-diagram}. The small lobe on the left side emerges at $\lambda = 0$, takes its maximum value at $\lambda \approx 0.1$, and disappears at $\lambda_\text{c1} = 0.295$, where all the interior spins are equally correlated with their neighbors. This dimer pattern (I) is weakly coupled to the boundary, and the order is almost independent of the boundary conditions even at small system size.  The large lobe on the right appears at $\lambda_\text{c1} = 0.295$, attains its maximum value at exactly $\lambda = 0.5$, and dies out at $\lambda_\text{c2} = 0.73$. For this pattern (II), the correlation function that defines the dimer profile is fully saturated ($\expectation{\sigma_i^z\sigma_{i+1}^z} = -1$ for odd $i$) at the $\lambda = 0.5$ peak.

\begin{figure}
\begin{center}
\includegraphics[width = 0.85\columnwidth]{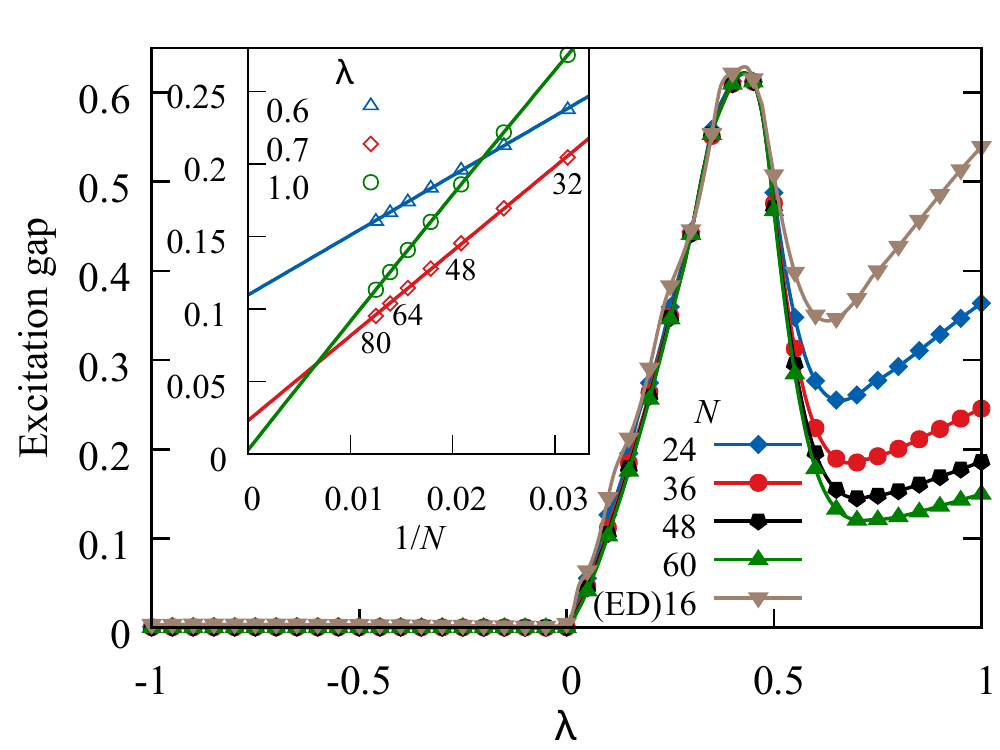}\\[0.2cm]
\includegraphics[width = 0.85\columnwidth]{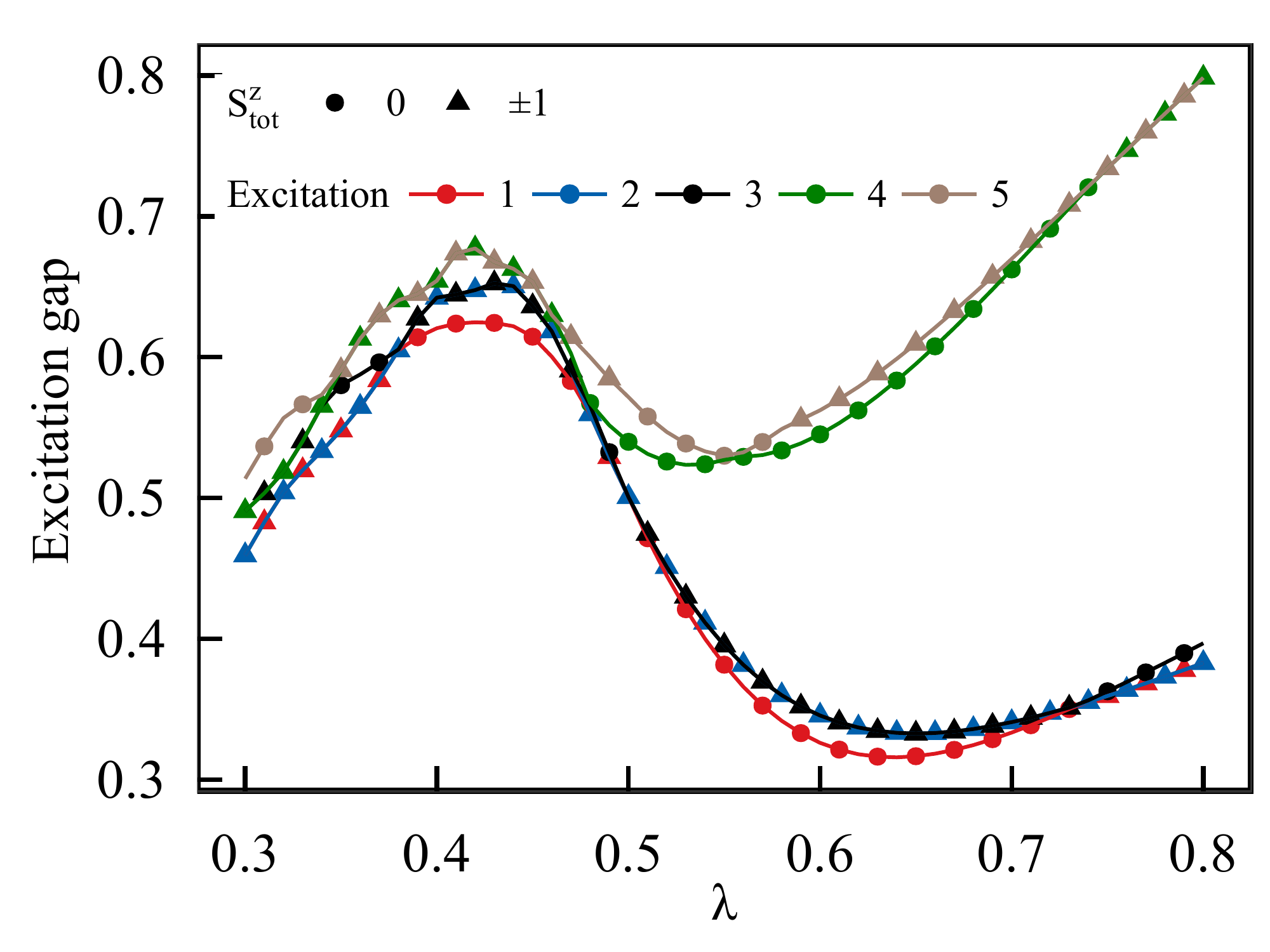}
\end{center}
\caption{\label{FIG:Gap}Top: The excitation gap reaches its maximum value ($\approx 0.62$) at $\lambda \approx 0.43$. The inset shows the excitation gap versus $1/N$, extrapolated to the thermodynamic limit. The system is clearly gapped for $\lambda = 0.6,0.7$ and the gapless at $\lambda = 1$, 
which is consistent with a phase transition at $\lambda_\text{c2} = 0.73$. In contrast, the excitation gap $\Delta \sim 1/N^z$ on the ferromagnetic side of the phase diagram scales with $z \approx 3$; cf.~Figs.~7 and B2 of Ref.~\onlinecite{Chen-JPA-17}.
Bottom: The data points shown represent the energy difference between each of the first five excited states and the ground state for the $N=18$ system over a range of $\lambda$ values in the vicinity of the pattern-II dimerized region. Colors indicate the ranking of the levels, and symbols denote the $S^z_\text{tot}$ character. Red circles show the $S^z_\text{tot} = 0$ states having lowest energy for $\lambda_\text{c1} < \lambda < \lambda_\text{c2}$. The next two levels (with $S^z_\text{tot} = \pm 1$) come close and touch at exactly $\lambda = 0.5$.
}
\end{figure}

Ising antiferromagnetism coexists with the dimerization, forming a dome of z-directed staggered spin order atop the two dimer-ordered lobes in Fig.~\ref{FIG:phase-diagram}. This gapped, doubly ordered region extends across $0 < \lambda < \lambda_\text{c2}$ and terminates with second-order phase transitions at either end. We expected to be able to extract
the critical exponents for these transitions from data collapse,
but the subleading corrections to finite-size scaling turn out to be
much too strong for us to do so reliably.

\begin{figure}
\begin{center}
\includegraphics[width = 0.85\columnwidth]{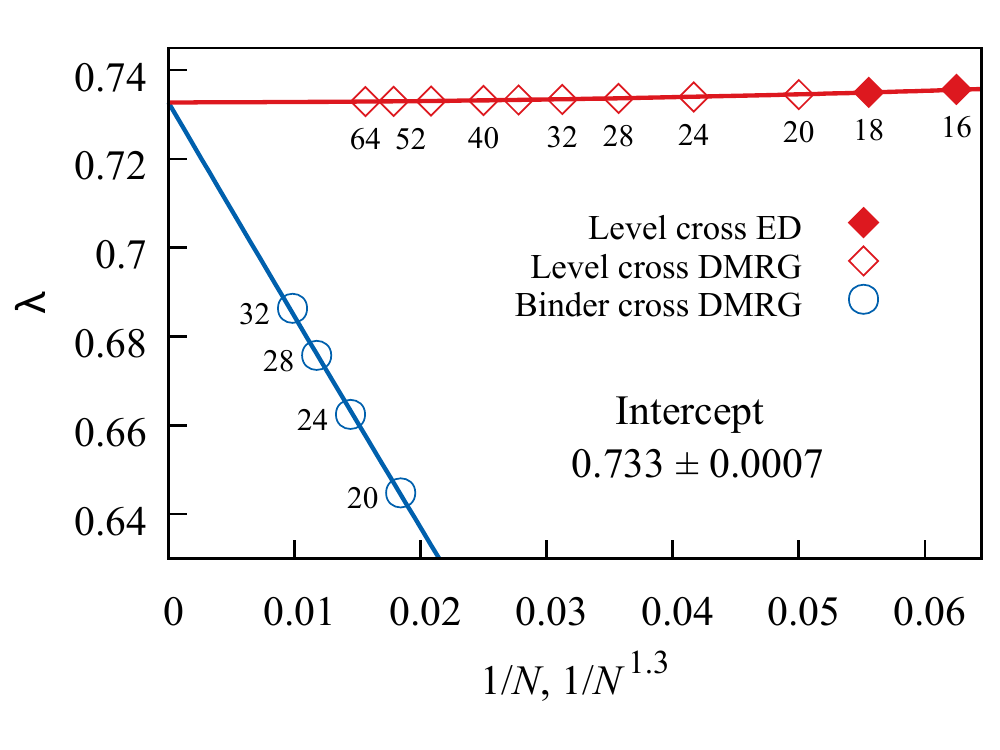}
\end{center}
\caption{\label{FIG:CPright} Shown here are the values of $\lambda$ at which cross (i) the $S^z_\text{tot} = 0$ and $S^z_\text{tot} = \pm 1$ first-excited-state energies and (ii) the Binder cumulants for doubled system sizes $N_1=N$ and $N_2=2N$. These are plotted with respect to $1/N$ and $1/N^{1/\nu}$ (using an estimated $\nu = 0.8$ for the correlation length exponent), respectively, and fit with second-order polynomials sharing a common y-axis intercept. The best common intercept is in agreement with our estimated critical point $\lambda_\text{c2} = 0.73$.
}
\end{figure}

We find that the ground state is unique and belongs to the $S_\text{tot}^z = 0$ sector for all values of $\lambda$. The lowest-lying excitations are doubly degenerate and $S^z_{\text{tot}}= \pm 1$ everywhere in the phase diagram except in the pattern II dimerized phase, where the excitation is a unique $S^z_{\text{tot}}=0$ state, and at $\lambda = 0.5$, where the $S_\text{tot}^z =0,\pm1$ excited states have the same energy. (It must be that those states are pinned together by symmetry, since the three-fold degeneracy holds regardless of system size.)
The excitation gap is plotted in Fig.~\ref{FIG:Gap}. Extrapolation from finite-system data to the $N \to \infty$ limit suggests that the system is gapped between $\lambda = 0^+$ and $\lambda = \lambda_\text{c2} = 0.73$, exactly coinciding with the region of dimer order and Ising antiferromagnetism. This is consistent with what we see in the Binder cumulant crossings and in the energy level crossings of the first excited states (as per Fig.~\ref{FIG:CPright}). 

\begin{figure}
\begin{center}
\includegraphics[width = 0.85\columnwidth]{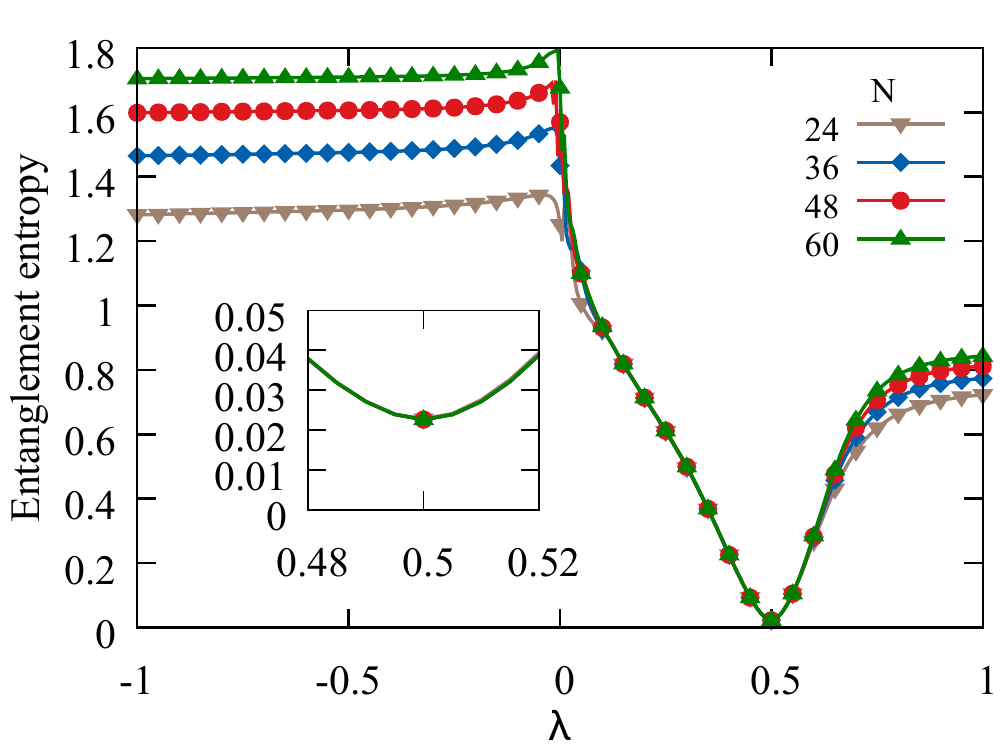}
\end{center}
\caption{\label{FIG:VnEntropy}The von Neumann entanglement energy for a bipartition of the spin chain is given as a function of $\lambda$. The peak value and maximum rate of growth with $N$ are located at $\lambda = 0$. For $\lambda = 0.5$, the entanglement entropy appears to be independent of system size, with no corrections to the area law~\cite{Hastings-JSATA-07}.
}
\end{figure}

\begin{figure*}
\begin{center}
\includegraphics{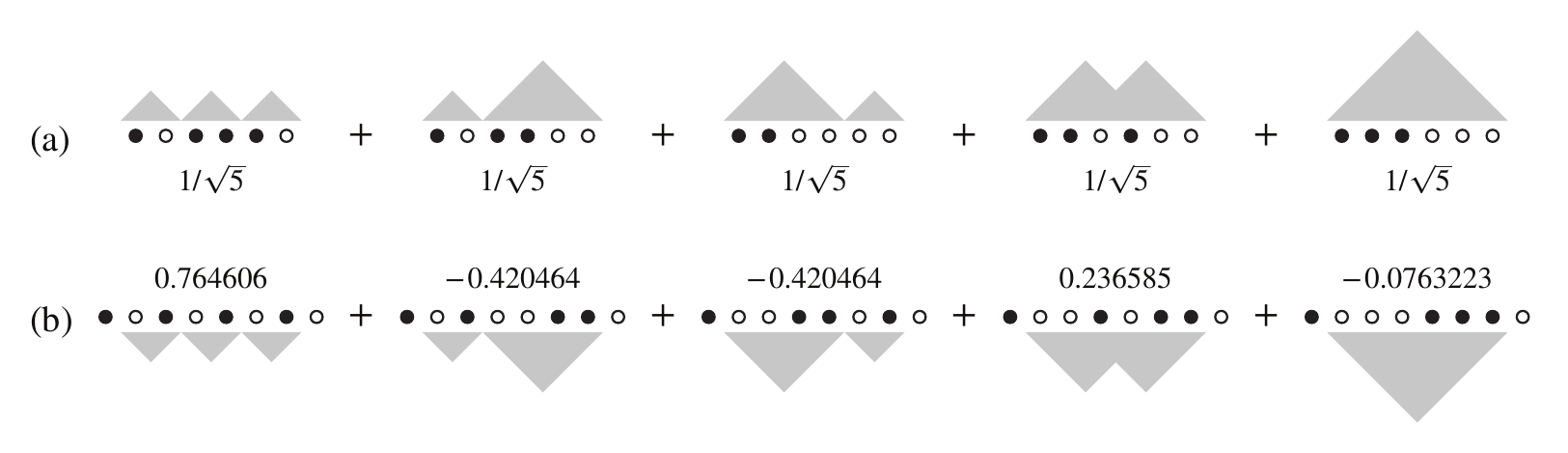}
\end{center}
\caption{\label{FIG:antiFredkin} 
Shown above are the particular superpositions of spin states that comprise 
the ground state wave functions for (a) $N=6$, $\lambda = 0$ 
and (b) $N=8$, $\lambda=1/2$. There is an exact correspondence between
the states contributing to the $N$-site chain at the Fredkin point
and those contributing to the ($N$+2)-site chain at the anti-Fredkin point.
At the Fredkin point, the only states contributing are those with a Dyck word structure---which is to say, those having a height profile $h_i = \sum_{i=1}^j \sigma_j^z$ that vanishes at the
end points ($h_0 = h_N = 0$) and never drops below the horizon at interior points ($h_i \ge 0$).
The structure at the anti-Fredkin point is a mirror-reflection landscape. 
A modified height function $\tilde{h}_i = \sum_{i=2}^j \sigma_j^z$ vanishes at the
end points ($\tilde{h}_1 = \tilde{h}_{N-1} = 0$) and never rises above the horizon ($\tilde{h}_i \le 0$).
This is a consequence of the re-emergence of the number of
Dyck word mismatches as a good quantum number for the system.
 }
\end{figure*}

Various features suggest that $\lambda = 0.5$ is an interesting special tuning point. We have already remarked that it is a point of maximum dimerization strength and that the excitations there are three-fold degenerate. It is also a point where the bipartite entanglement entropy is size-independent, as shown in Fig.~\ref{FIG:VnEntropy}. Direct investigation of the
ground state wave function itself reveals that the structure of the spin configurations that contribute at $\lambda = 0.5$ is similar in spirit to that obtained at $\lambda = 0$. 
Figure~\ref{FIG:antiFredkin} illustrates the connection using the language of the height profile. The underlying cause is that the number of Dyck work mismatches, appropriately defined, is again a good quantum number.

\section{Conclusions}\label{SECT:Conclusions}

The proposed model exhibits several interesting phase transitions and highly unusual scaling of the energy gap to the excited states. Figure~\ref{FIG:phase-diagram} shows our best estimate of the zero-temperature quantum phase diagram, based on an extrapolation to the thermodynamic limit from our numerical results on finite-size systems. 

The ground state evolves smoothly out of the boundary-twisted ferromagnetic state at $\lambda = -1$, and the strength of the xy ferromagnetic correlations increases monotonically to its peak value as $\lambda$ approaches zero. But the Fredkin state is seemingly not robust: even an infinitesimal amount of antiferromagnetic frustration proves to be completely disruptive. At $\lambda = 0^+$ the ground state changes abruptly. The transition is somewhat unusual, with strong xy ferromagnetic correlations on one side of the transition and vanishing Ising antiferromagnetism and dimer order on the other.

As $\lambda$ increases from $0^+$, an excitation gap opens up. The Ising antiferromagnetism and dimer order coexist, with the latter switching between two complementary short-range tilings (labeled I and II in Fig.~\ref{FIG:phase-diagram}) at $\lambda_\text{c1} = 0.295$.
The nature of the excitations in the type-II dimer region is different from everywhere else ($S^z_\text{tot} = 0$ rather than $S^z_\text{tot} = \pm 1$). 
All order appears to vanish simultaneously at $\lambda_\text{c2} = 0.73$. A more precise determination of this rightmost critical value is stymied by strong finite size effects; strictly speaking, we cannot rule out that the Ising antiferromagnetism and type-II dimer order vanish in a closely spaced sequence of transitions, but this seems unlikely.

A most unexpected feature is the presence of another special tuning point at $\lambda = 1/2$, where the ground state wave function mimics (in reverse, for the $N-2$ interior spins) the Dyck word structure that characterizes the $\lambda = 0$ Fredkin point. Here, however, the state is not an equal-weight superposition. Moreover, it has dimer and staggered Ising order, and its bipartite entanglement entropy is size-independent, suggesting a short-range-entangled, nearly product state.


\begin{thebibliography}{99}
\bibitem{comment2}
Away from the chain edges, the terms in Eq.~\eqref{EQ:bulk_hamiltonian} can be
regrouped as $\sum_i(U_{i-1} + D_{i+2})P_{i,i+1}$; this suggests
interpreting the Hamiltonian as a correlated Heisenberg model in which the ferromagnetic exchange coupling is sensitive to the configuration of neighboring spins. In contrast to the case of the conventional Heisenberg model, the privileging of the z direction in operators $U$ and $D$ reduces the global spin symmetry from SU(2) to U(1).
\end{thebibliography}
\end{document}